\documentclass[twoside,slac_two]{revtex4}
\usepackage{natbib}
\usepackage{graphicx}
\usepackage{fancyhdr}
\usepackage{enumerate}
\usepackage{amssymb}
\usepackage{amsbsy}
\usepackage{verbatim}
\usepackage{color}
\setcitestyle{square,numbers}
\begin{document}
\title{Implications of Einstein-Weyl Causality on Quantum Mechanics}
%
\author{D. J. BenDaniel}
\affiliation{Cornell University, Ithaca NY, 14853}
\begin{abstract} A fundamental physical principle that has consequences for the topology of space-time is the principle of Einstein-Weyl causality. We show here that this may have implications on quantum mechanics, as well. Borchers and Sen have rigorously investigated the mathematical implications of Einstein-Weyl causality and shown the denumerable space-time $Q^2$ would be implied. They then imbedded this space in a non-denumerable space but were left with important philosophical paradoxes regarding the nature of the physical real line {E}, e.g., whether {E} = {R}, the real line of mathematics. Alternatively, their initial result could suggest a constructible foundation.  We have pursued such a program and find it indeed provides a dense, denumerable space-time and, moreover, an interesting connection with quantum mechanics. \end{abstract}
\maketitle
\thispagestyle{fancy}
A fundamental physical principle that has consequences for the topology of space-time is the principle of Einstein-Weyl causality.   Borchers and Sen have rigorously investigated the mathematical implications of Einstein-Weyl causality and shown the denumerable space-time $Q^2$ would be implied \citep {Borchers}. They then imbedded this space in a non-denumerable space but were left with important philosophical paradoxes regarding the nature of the physical real line {E}, e.g., whether {E} = {R}, the real line of mathematics. Alternatively, their initial result could suggest a constructible foundation.  We have pursued such a program and find it indeed provides a dense, denumerable space-time and, moreover, an interesting connection with quantum mechanics. This paper has three parts. We first introduce this constructible foundation and show it contains polynomial functions which are locally homeomorphic with a dense, denumerable metric space $R^*$ and are inherently quantized. Uniformly continuous functions can then be effectively obtained by computational iteration. Secondly, postulating a Lagrangian for fields in a compactified space-time, we obtain a general description of which the Schr\"odinger equation is a special case. Thirdly, from these results we can then find that this denumerable space-time is relational (in the sense that space is not infinitesimally small if and only if it contains a quantized field) and, since $Q^2$ is imbedded in $R^{*2}$, it directly fulfills the mathematical requirements for Einstein-Weyl causality. Therefore, the theory predicts that {E} = $R^*$ and quantum mechanics provides a possible empirical corroboration. Finally, we discuss other possible physical implications of these results.

We propose the axioms in Table 1. The formulae for these axioms are given in the appendix.
\begin{table}[h]
\begin{center}
\caption{\label{tbl-axioms} Axioms}
\begin{tabular}{|l|p{5cm}|}
\hline
Extensionality & Two sets with just the same members are equal.\\
\hline
Pairs & For every two sets, there is a set that contains just them.\\
\hline
Union & For every set of sets, there is a set with just all their members.\\
\hline
Infinity & There are infinite ordinals $\omega^*$ (i.e., sets are transitive and well-ordered by $\in$-relation).\\
\hline
Replacement & Replacing the members of a set one-for-one creates a set (i.e., bijective replacement).\\
\hline
Regularity & Every non-empty set has a minimal member (i.e. ``weak'' regularity).\\
\hline
Arithmetic & Four axioms for predecessor uniqueness, addition and multiplication.\\
\hline
$\omega^*$-Constructibility & The axiom of comprehension restricted to constructible sets. \\
\hline
\end{tabular}
\label{example_table}
\end{center}
\end{table}
The first six axioms are the set theory of Zermelo-Fraenkel (ZF) without the power set axiom and with the axiom schema of subsets (a.k.a., separation) deleted from the axioms of regularity and replacement. Arithmetic is contained in ZF but must be axiomatized here. Because of the deletion of the axiom schema of subsets, a minimal $\omega^*$, usually denoted by $\omega$ and called the set of all finite natural numbers, cannot be shown to exist in this theory; instead this set theory is uniformly dependent on $\omega^*$ and then all the finite as well as infinitely many infinite natural numbers are included in $\omega^*$. These infinite numbers are one-to-one with $\omega^*$; a finite natural number is any member of $\omega^*$ that is not infinite. All the sets of finite natural numbers are finite. 

We have a sub-theory of ZF.  These axioms assert, in effect, that sets all are constructible. By constructible sets we mean sets that are generated sequentially by some process, one after the other, so that the process well-orders the sets. Historically, G\"odel had shown that an axiom asserting that all sets are constructible can be consistently added to ZF \citep{Godel}, giving a theory usually called ZFC$^+$. No more than countably many constructible subsets of $\omega^*$ can be shown to exist in ZFC$^+$ \citep{Fraenkel}. This result will, of course, hold for the sub-theory ZFC$^+$ minus the axiom schema of subsets and the power set axiom and with a restricted form of the axiom of comprehension.  We shall refer to these axioms as T.

We now will show that this theory can contain rather interesting polynomial functions. Let $w$ be an infinite natural number. Recall the definition of ``rational numbers'' as the set of ratios, in ZF called $Q$, of any two members of the set $\omega$. In T, we can likewise, using the axiom of unions, establish for $w$ the set of ratios of any two of its members. This will become an ``enlargement'' of the rational numbers and we shall call this enlargement $Q^*$. Two members of $Q^*$ are called ``identical'' if their ratio is 1. We now employ the symbol ``$\equiv$'' for ``is identical to.'' Furthermore, an ``infinitesimal'' is a member of $Q^*$ ``equal'' to 0, i.e., letting $y$ signify the member and employing the symbol ``='' to signify equality, $y=0 \leftrightarrow \forall k [y < 1/k]$, where $k$ is a finite natural number. The reciprocal of an infinitesimal is ``infinite''. A member of $Q^*$ that is not an infinitesimal and not infinite is ``finite''. Obviously, $y \equiv0\rightarrow\ y=0$. The constructibility axiom now allows creation of a set of constructible subsets of $\omega^*$ and, in addition, provides a distance measure, giving a metric space $R^*$. The members of $R^*$ represent the binimals (i.e., binary decimals) forming a dense, denumerable space.

An \textbf{\emph{equality-preserving}} bijective map $\phi(x,u)$ between intervals $X$ and $U$ of $R^*$ in which $x \in X$ and $u \in U$ such that $\forall x_1, x_2, u_1, u_2 [\phi(x_1, u_1) \wedge \phi (x_2,u_2) \rightarrow (x_1 - x_2 = 0 \leftrightarrow u_1 - u_2 = 0)]$ creates pieces which are biunique and homeomorphic. Note that $U=0$ if and only if $X=0$, i.e., the piece is inherently relational.
 
We can now define functions on $R^*$. $u(x)$ is a function of $x\in R^*$ if and only if it is a constant or a finite sequence of continuously connected biunique pieces such that some derivative is a constant. Thus, if not constant, $u(x)$ is a polynomial of bounded variation, uniformly continuous, locally homeomorphic with $R^*$ and \emph {with range $u(x)\neq 0$ $\leftrightarrow$ domain $u(x) \neq 0$}.  (Note:  Since infinitesimals can be defined in T, derivatives and integrals of polynomials can be obtained in the usual way.)  On the other hand, those power series such as sin(x), for which no derivative is constant, do not formally exist in this theory but can always be approximated as closely as required for physics by a sum of polynomials of sufficiently high degree obtained by an iteration of:
\begin{equation}
\int_a^b \left[p \left(\frac{du}{dx}\right)^2 - qu^2\right] dx \equiv \lambda \int_a^b ru^2 dx
\end{equation}
where $\lambda$ is minimized subject to:
\begin{equation}
\int_a^b ru^2 dx \equiv \textrm{const}
\end{equation}
where:
\begin{equation}
a \neq b, \quad u\left( \frac{du}{dx} \right) \equiv 0
\end{equation}
at $a$ and $b$; $p$, $q$, and $r$ are functions of the variable $x$.
Letting $n$ denote the $n^{th}$ iteration, $\forall k \exists n [\lambda_{n-1}-\lambda_n < 1/k]$ where $k$ is a finite natural number. So, a polynomial such that, say, $1/k < 10^{-50}$ should be sufficient for physics as it is effectively a Sturm-Liouville ``eigenfunction''. These can be decomposed, since they are polynomials, into biunique ``irreducible eigenfunction pieces'' obeying the boundary conditions.
As a bridge to physics, let $x_1$ be space and $x_2$ be time. We now postulate the following integral expression for a one-dimensional string $\Psi = u_1(x_1)u_2(x_2)$:
\begin{equation}
\int \left[ \left( \frac{\partial \Psi}{\partial x_1}\right)^2 - \left( \frac{\partial \Psi}{\partial x_2}\right)^2\right] dx_1dx_2 \equiv 0
\end{equation}
The eigenvalues $\lambda_{1m}$ are determined by the spatial boundary conditions. For each eigenstate $m$, we can use this integral expression constrained by the indicial relation $\lambda_{1m} \equiv \lambda_{2m}$ to iterate the eigenfunctions $u_{1m}$ and $u_{2m}$.

A more general string in finitely many space-like and time-like dimensions can likewise be produced. Let $u_{\ell mi}(x_i)$ and $u_{\ell mj}(x_j)$ be eigenfunctions with non-negative eigenvalues $\lambda_{\ell mi}$ and $\lambda_{\ell mj}$ respectively.
We define a ``field'' as a sum of eigenstates:
\begin{equation}
\underline{\Psi}_m = \sum_\ell \Psi_{\ell m}\underline{i_\ell}, \Psi_{\ell m} = \textrm{C}\prod_i u_{\ell mi}\prod_j u_{\ell mj}
\end{equation}
with the postulate: \textbf{\emph{for every eigenstate $m$ the Lagrangian form for the field equations in a compactified space-time is identically 0}}.
Let $ds$ represent $\prod_i r_idx_i$ and $d\tau$ represent $\prod_j r_jdx_j$. Then for all $m$,
\begin{eqnarray}
\lefteqn{
\int \sum_{\ell i} \frac{1}{r_i} \left[P_{\ell mi} \left( \frac{\partial \Psi_{\ell m}}{\partial x_i} \right) ^2 - Q_{\ell mi} \Psi^2_{\ell m} \right] ds d\tau } \label{eq-nsm0}\\
& & {}- \int \sum_{\ell j} \frac{1}{r_j} \left[P_{\ell mj} \left( \frac{\partial \Psi_{\ell m}}{\partial x_j} \right) ^2 - Q_{\ell mj} \Psi^2_{\ell m} \right] ds d\tau \equiv 0 \nonumber
\end{eqnarray}
In this integral expression the $P$, $Q$, and $R$ can be functions of any of the $x_i$ and $x_j$, thus of any $\Psi_{\ell m}$ as well. 
This is a \textbf{\emph{nonlinear sigma model}}. As seen in the case of a one-dimensional string, these $\Psi_m$ can in principle be obtained by iterations constrained by an indicial relation, $\sum_{\ell i} \lambda_{\ell mi} \equiv \sum_{\ell j} \lambda_{\ell mj}$ for all $m$.
We see that the postulate asserts a fundamental identity of the magnitudes of the two components of the integral.
 
A proof in T that the sum over all the eigenstates of each component has only discrete values will now be shown. Let expressions (\ref{eq-nsm1}) and (\ref{eq-nsm2}) both be represented by $\alpha$, since they are identical:
\begin{equation}
\sum_{\ell mi} \int \frac{1}{r_i} \left[P_{\ell mi} \left( \frac{\partial \Psi_{\ell m}}{\partial x_i} \right) ^2 - Q_{\ell mi} \Psi^2_{\ell m} \right] ds d\tau
\label{eq-nsm1}
\end{equation}
\begin{equation}
\sum_{\ell mj} \int \frac{1}{r_j} \left[P_{\ell mj} \left( \frac{\partial \Psi_{\ell m}}{\partial x_j} \right) ^2 - Q_{\ell mj} \Psi^2_{\ell m} \right] ds d\tau
\label{eq-nsm2}
\end{equation}
\begin{enumerate}[I.]
\item We assume that $Q_{\ell mj}\equiv 0$, $P_{\ell mj}\neq 0$,  that domain$\Psi \neq 0$ and is all of space-time and that $\alpha(\Psi)$ is non-negative and closed to addition and to the absolute value of subtraction. 
\item Since $\Psi$ is a function on $R^{*n}$, we recall that, if $\neg$ range$\Psi \equiv 0$ then range$\Psi \neq 0$ $\leftrightarrow$ domain$\Psi \neq 0$.  Accordingly, we obtain: if range$\Psi \equiv 0$ then $\alpha(\Psi)\equiv 0$ and if $\neg$range$\Psi \equiv 0$ then $\alpha(\Psi)\neq 0$.
\item \textbf {\emph{Therefore $\alpha(\Psi)$ has only discrete values} }  $\alpha (\Psi ) \equiv n \kappa$, where $n$ is any integer and $\kappa$ is some finite unit which must be determined empirically.
\end{enumerate}

With this result and without any additional physical postulates, we can now obtain the Schr\"odinger equation from the nonlinear sigma model in one time-like dimension and finitely many space-like dimensions. We need to look at only the time term.
Let $\ell = 1,2$, $r_t=P_{1mt}=P_{2mt}=1$, $Q_{1mt}=Q_{2mt}=0$, $\tau=\omega_mt$ and we normalize $\Psi$ as follows:
\begin{equation}
\Psi_m = \sqrt{(C/2\pi )}\prod_i u_{im}(x_i)[u_{1m}(\tau ) + \textbf{i}\cdot u_{2m}(\tau )]
\end{equation}
where $\textbf{i}=\sqrt{-1}$ with 
\begin{equation}
\int\sum_m\prod_i u_{im}^2 ds (u_{1m}^2+u_{2m}^2) \equiv 1
\end{equation}
We can then employ:
\begin{equation}
\frac{du_{1m}}{d\tau} = -u_{2m} \quad \textrm{and} \quad \frac{du_{2m}}{d\tau} = u_{1m}
\end{equation}
or
\begin{equation}
\frac{du_{1m}}{d\tau} = u_{2m} \quad \textrm{and} \quad \frac{du_{2m}}{d\tau} = -u_{1m}
\end{equation}
%
%
For the minimal non-vanishing field, $\alpha$ has its least finite value $\kappa$. Thus, 
\begin{eqnarray}
& & (C/2\pi) \sum_m \oint \int \left[ \left( \frac{du_{1m}}{d\tau}\right) ^2 + \left( \frac{du_{2m}}{d\tau}\right) ^2 \right] \nonumber\\
& & {} \prod_i u_{im}^2(x_i) ds d\tau \equiv C \equiv \kappa
\end{eqnarray}
Substituting the Planck constant $h$ for $\kappa$, this can now be put into the familiar Lagrangian form for the time term in the Schr\"odinger equation,
\begin{equation}
\frac{h}{2\textbf{i}} \sum_m \oint \int \left[ \Psi^*_m \left( \frac{\partial \Psi_m}{\partial t} \right) - \left( \frac{\partial \Psi^*_m}{\partial t} \right) \Psi_m \right] ds dt
\end{equation}
Since the Schr\"odinger equation is well confirmed by experiment, this can be considered an empirical determination of $\kappa$. 
\smallskip
\noindent
We can now show a link between quantum theory and space-time.
\begin{enumerate}[I.]
\item Assume $\exists \Psi\neg$ range$\Psi\equiv 0$ and domain$\Psi$ is all of space-time. With this we have: all of space-time $\neq 0$ $\rightarrow$ $\exists \Psi$ domain$\Psi\neq0$. Since domain$\Psi\neq0$ $\leftrightarrow$ range$\Psi\neq 0$ and range$\Psi \neq 0$ $\rightarrow$ $\alpha(\Psi)\neq 0$ we obtain:  all of space-time $\neq 0 \rightarrow\exists \Psi $ $ \alpha(\Psi)\neq 0$
\item Also, if all of space-time = 0, the upper and lower limits of all the integrals in the computation of $\alpha(\Psi)$ are equal so that:  all of space-time = 0 $\rightarrow$ $\forall\Psi$ $\alpha(\Psi)=0$.
\item Therefore: all of space-time $\neq 0$ $\leftrightarrow$ $\exists \Psi$ $\alpha(\Psi)\neq 0$. 
\item Furthermore, since we have empirically shown $\alpha(\Psi)\equiv n h$, it follows that: 
all of space-time $\neq 0$ $\leftrightarrow$
$\exists \Psi$ $\alpha(\Psi) \geq h$. $\alpha(\Psi) \geq h$ is the Uncertainty Principle.
\end{enumerate} 
We have thus shown that this denumerable space-time is relational in the sense that space is not infinitesimally small if and only if it contains a quantized field.

\begin{itemize}
\item Returning to Einstein-Weyl causality, Borchers and Sen have rigorously investigated its mathematical implications, regarded as a 
partial order, for the underlying spaces.  This partial order was axiomatized by them and shown to admit $Q^2$ as an ordered space. They subsequently proved that, given certain topological assumptions, these spaces can be densely imbedded in spaces that have locally the structure of differentiable manifolds. They were left, however, with important philosophical paradoxes regarding the fundamental nature of the physical real line {E}, e.g., whether {E} = {R}, the real line of mathematics.
We have viewed their initial result, without any further assumptions, instead as an insight into a denumerable space-time.  This suggested an investigation into a constructible foundation T. We have here shown that T indeed provides a dense, denumerable metric space $R^*$ that can support polynomial functions and that eigenfunctions governing physical fields can then be effectively obtained by an iterative computation.  Since these fields are by definition locally homeomorphic with $R^*$ and $Q^2$ is imbedded in $R^{*2}$, this space directly fulfills the topological requirements for Einstein-Weyl causality. Thus the theory predicts that {E} = $R^*$. We suggest that quantum mechanics provides an empirical corroboration of this theoretical prediction.
\item Finally, the Schr\"odinger equation is obtained in this constructible theory without reference to the statistical interpretation of the wave function, which, it can be argued, may be inferred from the equation itself and a requirement that quantum mechanics will reduce to its classical limit. \citep{Gottfried}. Philosophically, this suggests that the Schr\"odinger equation could be considered conceptually cumulative with prior physics. If so, it would resolve a long-standing controversy. 
\end{itemize}
In addition, though we do not have the opportunity here to discuss these 
points, we note that:
\begin{itemize}
\item The proposed theory does not have impredicative sets. This possibly suggests that this foundation allows no physical antinomies. That can be intuitively satisfying since, were there physical antinomies, the universe would tear itself apart.
\item Dyson \citep{Dyson} argued that the QED perturbation series cannot converge to a limit without a catastrophically unstable vacuum state and hence the series must be divergent. However, in this constructible theory no series limit is reached and thus an unstable vacuum state is not created.
\item This theory may have some bearing on Wigner's metaphysical question regarding the apparent unreasonable effectiveness of mathematics in physics \citep{Wigner}. The Schr\"odinger equation and its statistical interpretation along with the denumerability of space-time and its relational nature all arise here as a direct consequence of an axiomatic approach similar to the foundation of a mathematical system.
\end{itemize}
\section*{Appendix: ZF - Subsets - Power Set + Constructibility + Arithmetic}
\noindent
\emph{Extensionality}. Two sets with just the same members are equal. $\forall x \forall y \left( \forall z \left( z \in x \leftrightarrow z \in y \right) \rightarrow x = y \right)$
\smallskip
\noindent
\emph{Pairs}. For every two sets, there is a set that contains just them. $\forall x \forall y \exists z (\forall ww \in z \leftrightarrow w = x \vee w = y)$
\smallskip
\noindent
\emph{Union}. For every set of sets, there is a set with just all their members. $\forall x \exists y \forall z (z \in y \leftrightarrow \exists u (z \in u \wedge u \in x))$
\smallskip
\noindent
\emph{Infinity}. There are infinite ordinals $\omega^*$ (i.e., sets are transitive and well-ordered by $\in$-relation). $\exists\omega^* (O \in \omega^* \wedge \forall x (x \in \omega^* \rightarrow x \cup \{x\} \in \omega^*))$
\smallskip
\noindent
\emph{Replacement}. Replacing members of a set one-for-one creates a set (i.e., ``bijective'' replacement). Let $\phi(x,y)$ a formula in which $x$ and $y$ are free,\\
$\forall z \forall x \in z\forall y (\phi(x,y) \wedge \forall u \in z \forall v(\phi(u,v) \rightarrow u = x \leftrightarrow y = v)) \rightarrow \exists r \forall t (t \in r \leftrightarrow \exists s \in z\phi(s,t))$
\smallskip
\noindent
\emph{Regularity}. Every non-empty set has a minimal member (i.e. ``weak'' regularity). $\forall x (\exists yy \in x \rightarrow \exists y (y \in x \wedge \forall z \lnot (z \in x \wedge z \in y)))$
\smallskip
\noindent
\emph{$\omega^*$-Constructibility}. The axiom of comprehension restricted to constructible sets. $\forall \omega^*\exists S[(\omega^*, O) \in S \wedge \forall y \neq 0 \forall z[(y,z) \in S \leftrightarrow ((y - m_y)\cup\ m_y,z\cup\{z\}) \in S]]$, where the minimal element of $y$ is $m_y$.
\smallskip

\noindent

The four formulae (a) to (d) below are contained in ZF but must be adjoined to T as axioms.
\smallskip
\noindent
Let $x' = x \cup \{x\}$
\begin{enumerate}[(a)]
\item $\forall x \in \omega^* (x \neq O \leftrightarrow \exists y \in \omega^*(y'=x))$
\item $\forall x \forall y (x'=y' \rightarrow x=y)$
\end{enumerate}
\smallskip
\noindent
Let $x$ and $y$ be members of $\omega^*$ and $[x,y]$ and $[[x,y],z]$ represent ordered pairs.
\begin{enumerate}[(a)]
\setcounter{enumi}{2}
\item $\exists A \forall x \in \omega^* \forall y \in \omega^* E!z\in\omega^*([[O,O],O] \in A \wedge [[x,y],z] \in A \rightarrow [[x,y'],z'] \in A \wedge [[x',y],z']\in A)$; addition: $x+y=z$
\item $\exists M \forall x\!\in\!\omega^* \forall y \in \omega^* E!z \in \omega^*([[O,O],O] \in M \wedge [[x,y],z] \in M
\rightarrow [[x,y'],z+x] \in M \wedge [[x',y],z+y]\in M)$; multiplication: $x\cdot y = z$
\end{enumerate}
\smallskip
\noindent
Define $[a,b]_r$ such that $[a_1,b]_r + [a_2,b]_r \equiv [a_1+a_2,b]_r \textrm{ and } [a_1,b_1]_r \equiv [a_2,b_2]_r \leftrightarrow a_1 \cdot b_2 \equiv a_2 \cdot b_1$. The extended set of rationals $Q^*$ is the set of such pairs for all $a$ and $b$ in $\omega^*$.

\begin{thebibliography}{99} 
\bibitem{Borchers} Borchers, H.J., Sen, R.N., Mathematical Implications of Einstein-Weyl Causality, Springer, Berlin Heidelberg, 2006.
\bibitem{Godel} G\"odel, K., The consistency of the axiom of choice and of the generalized continuum hypothesis. \emph{Annals of Math. Studies}, 1940.
\bibitem{Fraenkel} Fraenkel, A. A., Bar-Hillel, Y., Levy, A., Foundations of Set Theory. North Holland, Amsterdam, 1958.
\bibitem{Gottfried} Gottfried, K., Inferring the statistical interpretation of quantum mechanics from the classical limit. \emph{Nature}, 2000, 405.
\bibitem{Dyson} Dyson, F. J., Divergence of perturbation theory in quantum electrodynamics. \emph{Phys.Rev.}, 1952, 85.
\bibitem{Wigner} Wigner, E. P., The unreasonable effectiveness of mathematics in the natural sciences. \emph{Comm. Pure and Appl. Math.} 1960, 13.
\end{thebibliography}

\end{document}